# Species-specific differences in follicular antral sizes result from diffusion-based limitations on the thickness of the granulosa cell layer.

M. Bächler[1,§], D. Menshykau[1,2,§], Ch. De Geyter[3], D. Iber[1,2,*]

1  Department for Biosystems, Science, and Engineering (D-BSSE), ETH Zürich
2  Swiss Institute of Bioinformatics, Switzerland
3  Clinic of Gynecological Endocrinology and Reproductive Medicine, Women's Hospital, University of Basel, Switzerland
§  These authors contributed equally

*Corresponding Author and Contact Person for Reprint Requests:

Dagmar Iber

Department for Biosystems, Science, and Engineering (D-BSSE),

ETH Zurich

Mattenstraße 26

4058 Basel

Switzerland

Tel + 41 61 387 3210

Fax + 41 61 387 31 94

dagmar.iber@bsse.ethz.ch

# Abstract

The size of mature oocytes is similar across mammalian species, yet the size of ovarian follicles increases with species size, with some ovarian follicles reaching diameters more than 1000-fold the size of the enclosed oocyte. Here we show that the different follicular sizes can be explained with diffusion-based limitations on the thickness of the hormone-secreting granulosa layer. By analysing published data on human follicular growth and granulosa cell expansion during follicular maturation we find that the 4-fold increase of the antral follicle diameter is entirely driven by an increase in the follicular fluid volume, while the thickness of the surrounding granulosa layer remains constant at about 45±10 μm. Based on the measured kinetic constants, the model reveals that the observed fall in the gonadotropin concentration from peripheral blood circulation to the follicular antrum is a result of sequestration in the granulosa. The model further shows that as a result of sequestration, an increased granulosa thickness cannot substantially increase estradiol production but rather deprives the oocyte from gonadotropins. Larger animals (with a larger blood volume) require more estradiol as produced by the ovaries to downregulate FSH-secretion in the pituitary. Larger follicle diameters result in larger follicle surface areas for constant granulosa layer thickness. The reported increase in follicular surface area in larger species indeed correlates linearly both with species mass and with the predicted increase in estradiol output. In summary, we propose a structural role for the antrum in that it determines the volume of the granulosa layer and thus the level of estrogen production.

# Introduction

Ovarian follicle development has been studied for decades and has enabled major progress in animal breeding and in human reproductive medicine (Adams, et al., 2012, Richards and Pangas, 2010). The overall process appears to be similar across mammals: during menstruation, increased secretion of follicle stimulating hormone (FSH) in the pituitary promotes the recruitment of a new cohort of growing follicles in the ovaries. FSH and luteinizing hormone (LH) are both transported via the blood circulation to the theca (Richards and Pangas, 2010) (Figure 1A,B), and stimulate there the production of androgens, which are converted to oestrogens in the granulosa cell layer of the follicle (Bao and Garverick, 1998, Erickson, et al., 1979, Silva and Price, 2002). The oestrogens, in turn, enter the blood and down-regulate the production and secretion of FSH in the pituitary (Kumar, et al., 1997). As the FSH concentration in the blood circulation subsequently falls, an increasing number of follicles undergoes atresia, and only the one follicle that can best compensate the lower FSH signalling with LH signalling survives and becomes the dominant, ovulating follicle (Xu, et al., 1995).

While the size of mature oocytes is similar across different mammalian species, the size of ovarian follicles differs greatly (Table 1), with some ovarian follicles reaching diameters more than 1000-fold the size of the oocyte that develops inside (Gosden and TELFER, 1987). Larger follicles are mainly characterized by a larger size of the fluid-filled cavity, the antrum. No direct role of the antrum in follicular function has so far been identified and it is an open question why the size of the antrum (and thus that of the follicle) is so variable among species. The follicular fluid that fills the antral cavity contains water, electrolytes, serum proteins and high concentrations of steroid hormones secreted by the surrounding granulosa cells (Rodgers and Irving-Rodgers, 2010). The membrana granulosa of the follicle is avascular and can therefore only be reached by diffusion from the vascularized theca (Figure 1A,B). The oocyte together with the surrounding cumulus cells forms the COC. The COC resides eccentrically in the follicle and is attached to the granulosa cell layer.

The geometry, the timing of the maturation process, and the core regulatory network that controls the follicle maturation process have been defined (Richards and Pangas, 2010). Within the ovarian follicle, FSH and LH regulate a large number of target genes and proteins, which interact in a complex regulatory network (Gloaguen, et al., 2011). Within the follicle, the most important regulatory factors are steroid hormones (androgens and oestrogens) but also insulin-like hormones, as summarized in Figure 1C. The regulatory network is further complicated by the spatial restriction of many of the gene expression domains. Thus, many of the signalling components are produced only in parts of the follicle, with some diffusing and others being cell-bound within the tissue.

Computational models have the potential to integrate large amounts of published information and can be used to evaluate the consistency of available data. A number of computational models have previously been developed to analyse aspects of the follicle maturation process (for a review, see (Vetharaniam, et al., 2010), but only few models explore the processes within the follicle itself. We have recently developed a model for the spatiotemporal signaling dynamics of the core regulatory network during bovine folliculogenesis (Iber and Geyter, 2013). The model was solved on a 1D-domain and demonstrated that the observed bovine gene expression patterns can indeed result from these core regulatory interactions. The time-dependent 3D-geometry of the human follicle and the rate of cellular expansion have been measured in great detail during the maturation process (Gougeon, 1986, McNatty, 1981). Recently, the gene expression levels have also been measured in human follicles at various stages of their development (Jeppesen, et al., 2012). Given the availability of these quantitative datasets, it is now feasible to construct a 3D-computational model for the human follicle maturation process. To this end, we extended the bovine model to human folliculogenesis using published quantitative human data on follicle growth, cell expansion, hormone concentrations and gene expression kinetics. Based on detailed quantitative growth data, the granulosa layer retains a constant thickness of about 45 μm during follicular growth, even though granulosa cells proliferate strongly. Thus, the massive expansion of the follicle from a diameter of 5 mm to 2 cm within 10-14 days of development is driven entirely by an increase in the volume of the follicular fluid. Based on the measured kinetic constants, the model further predicts that the granulosa cell layer would sequester hormones and that this would result in a concentration difference between the serum and the follicular fluid. The predicted concentration differences have been noted previously (Stone, et al., 1988) and we confirmed in measurements that model predictions and experiments agree quantitatively for a range of serum concentrations. We subsequently used the validated model to test the impact of altering either the size of the follicle or the thickness of the granulosa cell layer, and we find that, because of the diffusion-limitations, oestradiol production can rise substantially only by increasing the size of the follicle and thus the surface of the granulosa cell layer. We propose the diffusion-limitations across the granulosa cell layer as the reason for the follicular expansion in larger animals.

## Material and Methods

*The model*

Folliculogenesis in cattle and humans bears great similarity (Aerts and Bols, 2010). We therefore based the model for human folliculogenesis on our previous work on bovine folliculogenesis (Iber and Geyter, 2013). The model is formulated as isotropic advection-reaction-diffusion equations for a compound $c_i$ with diffusion coefficient $D_i$ and external velocity field **u**

$$\partial_t c_i + \nabla(\mathbf{u} c_i) = D_i \nabla^2 c_i + \mathcal{R}(c_i) \qquad (1)$$

$\mathcal{R}(c_i)$ denotes the reaction terms that describe the regulatory interactions of the hormones FSH (*F*), LH (*L*) and their receptors, of androgens (*A*), oestradiol (*E*), and the oestradiol receptor for steroid-dependent signaling, as well as of IGF signaling (*I*) as shown in Figure 1C. The reaction terms $\mathcal{R}(c_i)$ are listed in the Supplementary Material. Much as in the bovine model we use zero flux boundary conditions for all hormones, receptors and their complexes. The parameter values are listed in Table S1 in the Supplementary Material.

*Simulations*

The PDEs were solved with finite element methods as implemented in COMSOL Multiphysics 4.3a. COMSOL Multiphysics is a well-established software package and several studies confirm that COMSOL provides accurate solutions to reaction-diffusion equations both on constant (Cutress, et al., 2010) and growing domains (Carin, 2006, Thummler and Weddemann, 2007, Weddemann and Thummler, 2008). Simulations on growing domains were carried out using the Arbitrary-Lagrangian Eulerian (ALE) Method. Details on how biological models in general, and reaction-diffusion-advection equations on a domain comprising several subdomains in particular are implemented in COMSOL have been described by us previously (Germann, et al., 2011, Menshykau and Iber, 2012). The accuracy of the calculation for the components with sharpest concentration profiles (FSHR and LHR) was 2% or higher, except for a narrow range of time where the accuracy was 8% or higher.

*Measurements of the concentrations of FSH, LH and hCG*

In eight women treated with exogenous gonadotropins for ovarian hyperstimulation in assisted reproduction the follicular fluid of the first ovarian follicle in the right and in the left ovary were collected together with serum for the measurement of the concentrations of FSH, LH and HCG, respectively. The gonadotropin levels both in the serum and in the follicular fluid were measured by quantitative determination using the Elecsys System, Roche Diagnostics, Rotkreuz, Switzerland. All partipants were informed about the rationale of the project and signed informed consent.

## Results

*The growth of human ovarian follicles*

Detailed quantitative data sets on the growth of human follicles and on the expansion of the different compartments are available. The thickness of the granulosa cell layer and the follicular volume are the most important compartments in our model and detailed knowledge of their expansion (Gougeon, 1986, McNatty, 1981) allows us to simulate the regulatory network on a 3-dimensional axisymmetric follicular domain (Figure 1B) with realistic tissue layer sizes rather than on an idealized, linearly expanding 1D-domain, as was the case in our previous bovine model (Iber and Geyter, 2013). The thickness of the theca and the size of the COC have not been reported over time. However, the exact thickness of the theca does not impact on our model predictions because the theca is vascularised by blood capillaries. Therefore, the hormone concentrations in this layer are considered to directly reflect the serum hormone concentrations and to not be affected significantly by reaction and diffusion processes in this layer. In line with previous reports we will use 100 μm for the thickness of the theca throughout follicular development (Singh and Adams, 2000). The diameter of the mature COC at ovulation is about 3 mm. While the COC is well known to expand during follicle maturation we have not found detailed size measurements at earlier stages, and we therefore assume that the diameter of the COC expands at a similar rate as that of the follicle.

The preovulatory phase starts at a follicle diameter of around 5 mm and ends after around 14 days at 25 mm (Gougeon, 1986). The growth rate remains grossly constant which implies $v_F = 1.74 \times 10^{-5}$ mm s$^{-1}$ as growth rate of the follicle diameter. Both the volume of the follicular fluid (Figure 2A) and the number of granulosa cells (Figure 2B) have been reported for the entire preovulatory phase (Gougeon, 1986, McNatty, 1981). The follicle is about spherical (Figure 1B,C) and we can therefore use the relation $V_{FF} = 4/3\,\pi\,r_{FF}^{3}$ to convert the follicular fluid volume $V_{FF}$ into the radius of the follicular fluid compartment, $r_{FF}$ (Figure 2C). We note that the radius and follicular volume obtained with such a spherical approximation are very close to the measured data (compare measurements (dots) and spherical approximation (solid lines) in Figure 2A,C). The volume of human ovarian granulosa cells has been determined as $V_C = 1140$ μm$^3$ (Dhar, et al., 1996). By multiplying the number of granulosa cells with the cell volume we obtain the volume of the granulosa layer. Using again the spherical shape of the follicle we note that the volume of the granulosa layer is $V_G = 4/3\ \pi\ (r_G^3 - r_{FF}^3)$ where $r_G = r_F + s_G$ is the combined radius of follicular fluid and granulosa layer and $s_G$ is the thickness of the granulosa layer. We can thus calculate the thickness $s_G$ of the granulosa layer at different follicle sizes and find that the granulosa thickness is 45±10 μm throughout follicular development (Figure 2D). Finally we can use the growth rate $v_F$ to relate the follicle diameter to developmental time (Figure 2E, F). Again we stress that the approximations fit the measured data very well as evident by comparing the measurements (dots) and spherical approximation (solid lines) in Figure 2A-F.

The conversion of the measured diameters and cell numbers and volumes permits us to approximate the radii of the different compartments in our simulations over time (Figure 2G). The growth kinetics are described by six parameter values, as listed in Table S1. The geometry of the follicle used for modeling the 3D-follicle (Figure 1B) has a rotational axis of symmetry going through the center of follicle and the center of COC (Figure 2H). It should be noted that only the follicular fluid and the COC layers expand at speed $v_F$; the thickness of the thecal and granulosa cell layers do not change over time (Figure 2D,F) and the endpoints are thus only shifted as the follicular fluid and COC expand together (Figure 2G). The final diameter of the cumulus oocyte complex is 3 mm in the model.

*A model for human folliculogenesis*

Building on our model for bovine folliculogenesis (Iber and Geyter, 2013), we next sought to develop a model for the regulatory network that controls the development of the dominant human ovarian follicle in the preovulatory phase, aiming at integrating the known core regulatory interactions into a consistent framework. We formulated the model as a set of isotropic advection-reaction-diffusion equations that describe the regulatory interactions of the hormones FSH (*F*), LH (*L*), androgens (*A*), and oestradiol (*E*) as well as of IGF signalling (*I*). The ligands can diffuse inside the entire follicle, while their receptors are restricted to cells, thus diffuse at much lower speeds (Table S1) and cannot diffuse between the different tissue layers or in the follicular fluid (Figure 1A,B). Previous studies have successfully described the *in vivo* distribution of diffusible signalling molecules with continuous reaction-diffusion equations on a domain with a length scale as small as 10 cells (Iber and Zeller, 2012), and we therefore expect that the ligands in our model can also be adequately described by continuous reaction-diffusion equations. The receptors are more of a concern as these are restricted to cells and their diffusion is thus limited by the cell boundaries in the tissue. We have previously noted that during the receptor half-life receptors can diffuse over distances of less than the diameter of one epithelial cell (Menshykau, et al., 2012). For simplicity, we therefore also use continuous reaction-diffusion equations for the receptors.

The modelled core network regulating the development of the follicle is shown in Figure 1C and the reaction terms are listed in the Supplementary Material. In the mathematical formulation we consider four types of reactions: complex formation at rate $k_{on}$ and complex dissolution at rate $k_{off}$, linear decay at rate $\delta c_i$, where $c_i$ refers to the concentration of the component $_i$, and production at a constant rate $\rho$ or at a modulated rate $\rho f(\sigma(c_i, K_i))$. Here the σ term indicates a Hill function

$$\sigma_i = \frac{c_i^n}{c_i^n + K_i^n},$$

with Hill constant $K_i$, which specifies the concentration of $c_i$ where half-maximal activity is observed, and Hill coefficient *n* which defines the steepness of the response; we are using n = 2 throughout. The Hill functions can be used in different combinations to either specify activating influences, or by using $1 - \sigma_i$ for inhibitory impacts of $c_i$. The regulatory interactions along with the evidence have been discussed previously (Iber and Geyter, 2013). In brief, the following regulatory interactions are included in the model: FSH and LH, secreted by the pituitary gland and transported through the blood, move through the endothelium and sub-endothelial basal lamina of the thecal vasculature into the thecal layer of the follicle (Rodgers and Irving-Rodgers, 2010) and bind their receptors to form FSH and LH receptor ligand complexes (Fig. 1C, arrows (abbreviated as A# in the following) 1 and 2). *FSH-receptor* is constitutively expressed in granulosa and COC and its expression can be induced by signalling interactions in all tissue layers. FSH- and LH-signalling reduces the stability of the FSH and LH receptor mRNA (A3, A4) (Nair, et al., 2002, Schwall and Erickson, 1984, Themmen, et al., 1991), supports IGF signalling (A5) and enhances the production of androgens (A6) (Bao and Garverick, 1998), as well as the production and activity of aromatase (A7) (Erickson, Wang and Hsueh, 1979, Silva and Price, 2002), the enzyme that catalyses androgens into oestradiol (A8). Production of androgens is restricted to the theca, while its conversion to oestradiol is restricted to granulosa and COC. Oestradiol binds to its receptor, forming an oestrogen receptor-ligand complex (A9). Oestrogen signalling enhances the production of aromatase (A10), as well as the production of the receptors for FSH (A11), LH (A12), and oestradiol (A13) (Couse, et al., 2005, Richards, 1975, Sharma, et al., 1999). IGF signalling is necessary for the gonadotropin-dependent expression of *aromatase* (A14) (Silva and Price, 2002), enhances the production of FSH and LH receptors (A15, A16) (Hirakawa, et al., 1999, Minegishi, et al., 2000) and reduces *oestrogen receptor beta* expression (A17). We aimed at developing a parsimonious model with the simplest possible set of regulatory interactions that reproduce the measurements. Aromatase is therefore not explicitly included because its activity can well be approximated as the direct result of IGF signalling and of the regulation through FSH and LH.

The reaction terms are identical to those used in the bovine model (Iber and Geyter, 2013) except for two new regulatory interactions that had to be added to the human model to account for a marked difference in the bovine and human gene expression data. Thus, in the human granulosa cells *FSH-* *receptor* expression is high in small follicles (6 mm) and subsequently decreases (Jeppesen, Kristensen, Nielsen, Humaidan, Dal Canto, Fadini, Schmidt, Ernst and Yding Andersen, 2012), whereas in bovine follicles *FSH-receptor* expression is lower and increases during follicular growth (Bao and Garverick, 1998, Xu, Garverick, Smith, Smith, Hamilton and Youngquist, 1995). To account for this difference we needed to introduce a previously neglected negative feedback of gonadotropin signalling on *FSH- and LH-receptor* expression (Figure 1C, A3,A4) and increase the rate of FSH/LH/oestradiol-independent *FSH-receptor* expression, as described in the Supplementary Material.

We initiate the model without any hormones, receptors and hormone-receptor complexes, because we want to study the mechanisms that result in the emergence of the characteristic gene expression patterns in the follicle. We note that other homogenous initial conditions within the physiological range do not affect our model predictions. The only exception is the initial concentration of the IGF-signalling complex in the theca, which is important to reproduce the early expression of *LH-receptors* in the theca.

*Model consistency with human data*

Even though the model has 35 kinetic parameters and 6 growth parameters it is very well constrained (Table S1). We discussed the data-based derivation of the growth parameters in the previous section. The first 28 kinetic parameters values in Table S1 are the same as in the bovine model and all but two of these parameter values have been directly measured, sometimes in several independent experiments; two parameter values were previously established in the bovine model based on gene expression data (Iber and Geyter, 2013). The remaining seven parameter values had to be changed to account for the differences in the measured gene expression time courses and steroid concentrations in women and cattle. Thus the androgen concentration is higher in human than in bovine follicles whereas the oestradiol concentration is lower (Bao and Garverick, 1998, Jeppesen, Kristensen, Nielsen, Humaidan, Dal Canto, Fadini, Schmidt, Ernst and Yding Andersen, 2012, Xu, Garverick,

Smith, Smith, Hamilton and Youngquist, 1995). The reported human concentration ranges are shown in Fig. 3A as shadings. To reproduce the human follicular androgen concentration the androgen production rate $\rho_A$ had to be increased fourfold. To reproduce the measured human oestrogen concentration the Hill constant for IGF-dependent regulatory processes, $K_I$, had to be lowered. The *FSH-receptor* expression has been found to decrease over time (Jeppesen, Kristensen, Nielsen, Humaidan, Dal Canto, Fadini, Schmidt, Ernst and Yding Andersen, 2012). The LH and FSH receptor mRNAs have previously been reported to be destabilized in response to gonadotropin receptor dependent signalling (Nair, Kash, Peegel and Menon, 2002, Schwall and Erickson, 1984, Themmen, Blok, Post, Baarends, Hoogerbrugge, Parmentier, Vassart and Grootegoed). To achieve such a negative feedback (Figure 1C, arrows 3,4) the signalling threshold $K_F$ had to be lowered (Figure 3B). Since both FSH and LH-receptors are G-protein-coupled receptors that link to the same signalling machinery (Richards, et al., 2002, Wood and Strauss) we now use equal thresholds for FSH and LH signalling, i.e. $K_L = K_F$. To still obtain similar *LH-receptor* expression levels we had to increase the production rate for the IGF receptor complex, $\rho_I$, some 5-fold. Finally, there is an FSH/LH/oestrogen-independent regulation of *FSH-receptor* expression in granulosa and COC (Couse, Yates, Deroo and Korach, 2005, Zhou, et al., 1997), which we assume has a constant activity $\vartheta_G$ and $\vartheta_{COC}$. This value had to be increased in the granulosa to achieve a high initial rate of *FSH-receptor* expression that can then decrease as a result of the negative feedback (Figure 3B). The activity in the COC, $\vartheta_{COC}$, had to be lowered to limit the total receptor concentration to a physiologically realistic range. With these changes, the measured gene expression rates for *FSH-* and *LH- receptors* as well as for aromatase are well reproduced (Figure 3B) and the total LH receptor concentration remains below 3 nM (Figure 3C) as measured in experiments (Erickson, Wang and Hsueh, 1979).

*Spatiotemporal dynamics of the regulatory network in the follicle*

With the model and parameters all set we first simulated the physiological situation of follicle maturation in healthy women. The human serum levels of FSH and LH change during the menstrual cycle (Brindle, et al., 2006) and we used the measured FSH ad LH concentrations as the thecal gonadotropin concentrations (Figure 4A,B). The simulations show how the hormones enter the follicle from the theca by diffusion, and how the concentration increases slowly inside the follicle (Figure 4C,D). Interestingly, also the final hormone concentrations in the follicular fluid are much lower than in the serum (Figure 4C,D, solid lines). This is the result of hormone sequestration in the membrana granulosa by receptor binding (Figure 4E,F). We notice that most FSH is bound in the granulosa layer (Figure 4E, white layer), while most LH is bound in the theca (Figure 4F, dark grey layer), and some less in the granulosa (Figure 4F, white layer). These distributions are also in good agreement with the reported expression patterns of *FSH-* and *LH-receptors* (Figure 4G,H). Thus *FSH-receptors* are mainly expressed in the membrane granulosa (Rhind, et al., 1992), while *LH-receptors* are initially expressed mainly in the theca and later also in the granulosa (Bao and Garverick, 1998, Xu, Garverick, Smith, Smith, Hamilton and Youngquist, 1995). As a result of hormone sequestration little FSH and LH reaches the COC (Figure 4E,F, lightly shaded layer) and little receptor is expressed, which agrees well with the measured low receptor expression rates in the COC (Assou, et al., 2006, Eppig, et al., 1997, Jeppesen, Kristensen, Nielsen, Humaidan, Dal Canto, Fadini, Schmidt, Ernst and Yding Andersen, 2012). In the follicular fluid, hormones have a uniform concentration because they are mixed rapidly as compared to the timescale of development due to convection with moving fluid (Figure 4C,D, grey-shaded area).

*Sensitivity to parameter values and initial conditions*

While the parameter values are all based on experimental measurements (Table S1), these measurements were carried out in a range of different systems and may suffer from experimental errors and inaccuracies. We therefore checked the robustness of the observed concentration difference between serum and follicular fluid to changes in the parameter values. To that end we re-simulated our model with parameter values that were drawn from a Gaussian distribution with mean value as given in Table S1 and relative standard deviation 0.1. The standard deviation of the simulations at the final time point (day 14, depicted as grey shadow in Figure 4C-F) demonstrates little impact of such deviations in the parameter values on the relative FSH and LH concentrations between serum and follicular fluid (Figure 4C,D). We, however, noted considerable variance in the extent to which FSH- or LH-bound receptor complexes emerge in the theca (Figure 4E,F).

To identify the parameters with the largest impact on the relative gonadotropin concentration in follicular fluid and serum we carried out a sensitivity analysis. Here we calculated the relative change in the FSH concentration ratio (referred to as Ratio) in the two compartments in response to a 1% change in the parameter values $p_i$, i.e.

$$S = \frac{\Delta Ratio/Ratio}{\Delta p_i/p_i}.$$

Figure 5 includes all parameters for which $S > 0.001$. We note that the largest impact is observed for the gonadotropin diffusion coefficient $D_H$, the FSH receptor production rates $\rho_{FR}, \rho_{FRG,}$ and $\rho_{FRcoc}$, the decay rate of the FSH-receptor complex $\delta_{FRc}$, and the FSH response threshold $K_{FR}$. This further supports the observation that the gonadotropin concentration difference between serum and follicular fluid results from receptor-dependent sequestration in the theca and granulosa (Figure 4E,F). The initial conditions on the other hand have little impact, as long as they are homogenous in space.

*Concentration differences between the serum and the follicular fluid as a result of hormone sequestration in the granulosa cell layer*

The predicted concentration differences between serum and follicular fluid are evident already in 25-year old published data (Stone, Serafini, Batzofin, Quinn, Kerin and Marrs, 1988). However, since these concentration differences have received no prior attention and are central to the predictions of the present model, we sought to confirm these concentration differences between serum and follicular fluid by measuring the FSH, LH, and hCG concentrations in women. Thus, we assayed the hormone concentrations in the first aspirated left and right follicle of eight infertile patients undergoing oocyte retrieval for assisted reproduction and in their serum. These patients were treated with exogenous human menopausal gonadotropins or with recombinant FSH and had received one single subcutaneous bolus injection of human urinary chorionic gonadotropin (hCG) for ovulation induction, a substitute for LH with a longer serum half-life, but with a receptor-dependent turn-over rate of hCG similar to that of LH (Nakamura, et al., 2000). To compare the measured values to the simulations we converted the reported IU/l units into SI units using the following conversion factors: 1 IU/l FSH = 1.5 nM FSH, 1 IU/l LH = 0.75 nM (Olivares, et al., 2000) and 1IU/l hCG = 2.88 pM hCG (Stenman, 2004). We subsequently combined the measured LH and hCG concentrations as both proteins bind to the LH receptor with comparable affinity. As predicted by the model we find a much lower FSH and LH/hCG concentration in the follicular fluid compared to the serum (Figure 6A,B). As predicted by the model we find much lower concentrations of FSH, LH and hCG in the follicular fluid as compared to the serum (Figure 6A,B). During ovarian hyperstimulation the patients received different daily doses of FSH and LH, and we note that the differences between serum and follicular fluid concentrations decreases as the serum gonadotropin levels increase (Figure 6A,B). We checked whether we would obtain a similar saturation effect in the model, and we indeed obtain a similar result in the model (Figure 6C,D). For ease of comparison we include the line that best fitted the experimental data in the simulation plots. This line aligns well with the simulated predictions (Figure 6C,D).

*Diffusion-based limits in granulosa cell layer thickness can explain scaling of follicle size with species weight*

It has long been noticed that the size of follicles scales with the average species size (Table 1, Figure 7A) (Gosden and TELFER, 1987). It has been argued that this serves to increase the volume of the granulosa, and thus hormone production (Gosden and TELFER, 1987), but it has remained unclear why the volume of the granulosa would be increased via an expansion of the fluid-filled antrum rather than by a thickening of the granulosa layer. We used our parameterized model to explore the impact of increasing either the size of the fluid-filled antrum (as observed in nature) or the thickness of the granulosa on the oestradiol production. When we analyse the oestradiol production during follicle development relative to the surface area of the follicle (while keeping the thickness of the granulosa and the diameter of the COC constant) we observe a linear relationship (Figure 7B, slope 0.991+/- 0.002, $R^2$=0.99996), much as is observed in the differently sized species (Figure 7A, slope 1.02 +/- 0.08, R2=0.885). Assuming that heavier animals require more oestrogens because their blood volume increases in parallel to body mass, the present analysis suggests that larger follicles can provide the required higher oestradiol levels by accommodating more granulosa cells in the granulosa layer lining the follicular wall.

We next analysed the alternative option of increasing the thickness of the granulosa cell layer. We find, that even if the thickness of the granulosa is strongly increased (while keeping the diameters of the follicle and of the COC constant), the total production of oestradial rises by at most threefold., presumably because, as the thickness of the granulosa cell layer increases, insufficient gonadotropins and androgen precursors reach the inner part of the granulosa layer, thereby limiting the activity of aromatase. A tenfold decrease in granulosa thickness on the other hand, reduces the oestradiol concentration to almost zero (Figure 7C). A further limitation in altering oestradiol production via the granulosa thickness results from the follicular gonadotropin concentration: a fourfold increase in the thickness of the granulosa layer reduces the intrafollicular gonadotropin concentration to almost zero, while a reduction in granulosa thickness doubles the follicular gonadotropin concentration (Figure 7C). There is therefore little room for the up-regulation of the oestradiol production by increasing the granulosa thickness without affecting other processes. In summary, larger animals require the production of more oestradiol to achieve the same oestradiol serum concentration despite their higher blood circulation volumes, but this necessity cannot be accommodated by increasing the thickness of the granulose cell layer. Instead, their follicles need to become larger to accommodate more granulosa cells in a layer with similar thickness.

Gonadotropins are sequestered by their receptors, and polymorphisms in the α-oestrogen and the FSH-receptor genes have been associated with infertility (M'rabet, et al., 2012, Perez Mayorga, et al., 2000). The detailed biochemical impact of the changes in the FSH-receptor gene is not known but it has previously been noticed that reduction in binding capacity (NOT affinity) of LH- and FSH-receptor results in severe defects (Aittomäki, et al., 1995). While changes in the *LH-* and *FSH-receptor* expression levels have a much milder effect in the model than altering the granulosa cell layer thickness, a tenfold change in receptor expression rates still raises the levels of FSH and of oestradiol (but not of LH) in the follicular fluid by about twofold (Figure 7D), thus potentially providing a mechanistic explanation for the observed clinical effects of the polymorphisms.

Finally, we note that variations in granulosa cell layer thickness within a species are likely to lead to disease as the hormone balance will be altered and the system is not self-correcting. Thus, a thicker granulosa layer will result in a reduced intrafollicular FSH concentration and a higher oestradiol concentration; the higher oestradiol concentration will down-regulate the release of FSH from the pituitary, thus further reducing the serum FSH-concentration (Figure 7C). Infertile women are often treated with exogenous FSH to stimulate ovarian function thereby increasing the proliferation of granulosa cells and circulating levels of oestradiol. We therefore tested whether the observed defects could be alleviated by adding FSH. To that end we increased and decreased the granulosa cell layer width and FSH-concentrations in parallel (Figure 7E). Raising the FSH-concentration in the serum compensates only for minor thickening of the granulosa. As the granulosa cell layer reaches 7-fold its normal thickness FSH may no longer reach the follicular fluid and therefore not reach the enclosed COC, even if additional external FSH is applied. In contrast, compensation by exogenous gonadotropins would be successful in case of increased FSH-receptor expression levels (Figure 7D.).

## Discussion

Growth of the ovarian follicle is mainly driven by the expansion of a fluid-filled cavity, the antrum. Follicular size is a key marker of successful follicular development, and sonographic measurement of follicular size is the single most reliable parameter for clinical decision making in ovarian hyperstimulation. Despite its eminent importance in the monitoring of ovarian follicular growth, no direct role of the antrum has, however, so far been identified, and it is also an open question why the size of the antrum (and thus that of the follicle) differs to such an extent among different species.

By analysing published data on granulosa cell expansion and by computing the 4D-spatiotemporal events during follicular growth we demonstrate that the massive expansion of the follicle from a diameter of 5 mm to 2 cm within 10-14 days is entirely driven by an increase of the volume of the follicular fluid, while the thickness of the granulosa cell layer remains constant at about 45±10 µm, even though granulosa cells strongly proliferate (Figure 2). A similar thickness (~50 µm) of the granulosa cell layer has also been observed in other species, i.e. bovine (van Wezel, et al., 1999), goat and sheep (Mohammadpour, 2007). Based on the measured kinetic constants we predict that granulosa cells sequester gonadotropins as they diffuse from the peripheral blood circulation into the follicular antrum (Figure 4,5) and we confirm the existence of a FSH- and LH-concentration gradient between the serum and the follicular fluid (Figure 6). Based on these calculations, we conclude that a thicker granulosa cell layer would sequester and deplete important signalling factors, including FSH, delivered from the blood circulation towards the COC (Figure 7).

The fluid-filled follicular antrum determines the surface area of the follicle and thereby the volume of the oestrogen-secreting granulosa cell layer. In mono-ovulatory species, the appropriate serum concentration of oestrogens is crucial for the selection of the dominant follicle through the fine-tuning of the FSH-secretion in the pituitary (Xu, Garverick, Smith, Smith, Hamilton and Youngquist, 1995). Due to their copious circulating blood volume, larger animals require the production of more oestrogen and hence more granulosa cells to achieve the same modulatory effects in the pituitary. Increasing the size of the antrum in large animals allows for the expansion of the volume of the granulosa cell layer (and thus the production of more oestrogens) without increasing the thickness of the granulosa cell layer.

In bovine follicles a thickness range of 40-100 µm has been reported for small antral follicles (diameter ~5 mm), and a much narrower range (close to 50 µm) for later stages of follicular development (van Wezel, Krupa and Rodgers, 1999). The observed variability may result from differences in cell shape and from an uncorrelated proliferation of granulosa cells and antrum expansion (Rodgers, et al., 2001). According to our model, a narrow range of thickness of the granulosa cell layer is important, because a thin granulosa cell layer would fail to produce sufficient amounts of oestrogens to downregulate FSH secretion in the pituitary, while a thicker granulosa cell layer, while producing more oestradiol, would result in insufficient intrafollicular LH and FSH signalling due to limited diffusion (Figure 7). In addition, higher oestrogen levels would potentially prematurely lower the secretion of FSH by the pituitary and thereby additionally compromise the development of the oocyte in the COC. The small variability of granulosa cell layer thickness during the later stages of follicular development may thus derive from the sensitive impact of the oestradiol-producing granulosa cell layer on the delicate balance between the acquisition of follicular dominance and atresia of follicles.

In assisted reproductive medicine, the addition of exogenous hormones during ovarian hyperstimulation not only increases the number of growing follicles but also raises the oestrogen levels per mature follicle and the proliferation rate of the granulosa cells, in particular in those cases with high follicle numbers (Attaran, et al., 1998, Chanchal Gupta, 2012, De Geyter, et al., 1992). As a result, exaggerate thickening of the mural granulosa cell layer may block the diffusion of gonadotropins and other hormones to the COC. During final follicular development, these constraints may be compensated to some extent by the increased formation of perifollicular capillaries. Interestingly, increased density of blood capillaries in the theca after wedge resection has been shown to revert the ovulatory function of polycystic ovaries due to enhanced delivery of FSH to the granulosa (Inzunza, et al., 2007). Moreover, earlier reports show that increasing the doses of gonadotropins during ovarian hyperstimulation (step-up regimen) results in more collected oocytes (Christin-Maitre et al., 2003), while lower numbers of oocytes are collected if the administration of FSH is withheld at the end of follicular development, as in prolonged coasting (D'Angelo et al., 2011).

Data-based, validated computational models of biomedical processes are still rare, but they are likely to become invaluable tools to define the molecular causes of disease and to develop novel and individual therapeutic approaches that respect the complex regulatory logic of biological systems. Currently, in clinical reproductive medicine, the choice of the daily FSH-dosage to be administered is exclusively based on a quantitative assessment of ovarian reserve, as given by the antral follicle count or the concentration of the anti-Muellerian hormone in the serum, but not on the FSH-receptor density or the capacity of the granulosa cells to proliferate. Mathematical models encompassing individual differences in receptor densities and activity may help to understand these effects in individual patients during ovarian hyperstimulation and perhaps assist in designing appropriate treatment modalities in each case prospectively.

## Author contributions

DI and CDG designed the study; MB, DM, and DI carried out the analysis; DI, CDG, DM wrote the paper; all authors approved the final manuscript.

## Funding

This work was supported by the Repronatal Foundation, Basel Switzerland.

# TABLES

Table I: The diameter of the mature follicle and characteristic weight in different mammalian species.

| Follicle [mm] | Weight [kg] | Specie | Ref |
|---|---|---|---|
| 0.42 | 0.03 | Mouse | (Griffin, et al., 2006) |
| 0.55 | 0.2-0.25 | Albino Rat | (SANGHA and GURAYA, 1989) |
| 0.64 | 0.2 | Hamster | (Griffin, Emery, Huang, Peterson and Carrell, 2006) |
| 2.8 | 2.7 | Rabbit | (Osteen and Mills, 1980) |
| 4 | 3 | Cat | (Izumi, et al., 2012) |
| 6 | 9-10 | Beagle (Dog) | (Reynaud, et al., 2009) |
| 6 | 23 | Sheep | (Aurich, 2011) |
| 7.5 | 35 | Serrana Goat | (Simões, et al., 2006) |
| 8 | 150 | Gilt | (Chiou, et al., 2004) |
| 7-12 | 48-84 | Alpaca | (Bravo, et al., 1991) |
| 7-12 | 130-200 | Llama | (Bravo, Stabenfeldt, Lasley and Fowler, 1991) |
| 16 | 250-800 | Water Buffalo | (Taneja, et al., 1996) |
| 20 | 700 | Cow | (Evans, 2003) |
| 20 | 2700 | Elephant | (Lueders, et al., 2011) |
| 23 | 60 | Human | (Evans, 2003) |
| 20-25 | 800 | Summatran rhinoceros | (Hermes, et al., 2007) |
| 27-38 | 300-550 | Camel dromedarius | (Manjunatha, et al., 2012) |
| 30-34 | 1800 | White rhinoceros | (Hermes, Göritz and Streich, 2007) |
| 50 | 1000 | Black rhinoceros | (Hermes, Göritz and Streich, |

|  |  |  | 2007) |
|---|---|---|---|
| 55 | 450 | Horse | (Aurich, 2011) |
| 120 | 1900 | Indian rhinoceros | (Hermes, Göritz and Streich, 2007) |

## Legends of the Figures

**Figure 1** - (A) A schematic 2D representation of an ovarian follicle. The follicle is a multilayered structure. Inside the follicle is the fluid-filled antrum, which is surrounded by a granulosa cell layer. The outer layer, the theca, is surrounded by a mesh of capillary blood vessels. All other parts of the follicle are avascular. The oocyte together with the surrounding cumulus cells forms the COC. The COC lies on one side of the follicle and is attached to the granulosa cell layer. (B) The 3D computational domain for the follicle. (C) The modeled signaling network for the regulation of follicular development, including FSH, LH, oestrogens [E], androgens [A], and IGF [I] signaling. Receptors and ligand-receptor complexes of component j are indicated as $R_j$ and $C_j$ respectively. Black dotted arrows indicate exchange with the blood, black solid arrows indicate chemical reactions (binding or catalysis), light gray arrows indicate activating impacts and arrows in darker grey indicate inhibitory impacts. All components also decay, but for greater clarity decay reactions have not been included in the scheme. For a more detailed discussion of the reaction network along with the evidence see the main text; numbers in brackets refer to single reactions as called out in the main text.

**Figure 2** - Follicular growth. (A) The volume of the follicular fluid at different follicle diameters $d_F$. The data (dots) was reproduced from [12]. The line shows the follicular fluid volume obtained with $V_{FF} = 4/3\,\pi\,r_{FF}^3$ where $r_{FF}$ was obtained by fitting the data points. (B) The number of granulosa cells in human Graafian follicles at different follicle diameters $d_F$. The data (dots) was reproduced from [12]. The light grey solid line shows the number of granulosa cells obtained by fitting the data points. The dark grey solid line shows number of granulosa cells obtained when converting a constant granulosa thickness $s_G = r_G - r_{FF} = 44\ \mu m$ into the number of granulosa cells $N_G = V_G/V_C$, where $V_G = 4/3\,\pi\,(r_G^3 - r_{FF}^3)$ is the granulosa layer volume and $V_C = 1140\ \mu m^3$ is the measured granulosa cell volume. (C) The radius of the follicular fluid as obtained by fitting the volume data in panel (A) using the relation $V_{FF} = 4/3\,\pi\,r_{FF}^3$. (D) The granulosa layer thickness $s_G$ in human graafian follicles at different follicle

diameters. The thickess was calculated by determining the granulosa volume $V_G = N_G \times V_C$ as the product of granulosa cell number $N_G$ and cell volume $V_C = 1140 \, \mu m^3$. The granulosa thickness $s_G = r_G - r_{FF}$ was then obtained from $V_G = 4/3 \pi s_G^3$. (E) The radius of the follicular fluid and (F) the thickness of the granulosa layer versus developmental time as obtained by converting follicle diameter in panels (C) and (D) using $d_F = 5mm + v_F t$. (G) The radii of the different compartments over simulation time. The radius $r_{FF}$ represents the radius of the fluid filled cavity (dotted dark-grey line). The radius $r_{COC} = r_{FF} - d_{COC}$ defines the start of the COC domain (black dash-dot line); $d_{COC}$ is the diameter of the cumulus oocyte complex. The radius of the sphere that includes both follicular fluid and granulosa cells layer has radius $r_G = s_G + r_{FF}$ (light grey dashed line), where $s_G$ is the thickness of the granulosa layer. The outer delimeter of the follicle is given by the radius $r_T = r_G + s_T$ (black solid line), where $s_T$ is the thickness of the the thecal tissue. (H) A section along the axis of symmetry of the computational domain.

**Figure 3** –Model Consistency with human data. (A) The measured (shaded areas) and simulated (lines) steroid concentrations of androgens (dash-dotted line, darker shading) and oestradiol (solid line, lighter shading) in the follicular fluid over time. The data was recorded by (Jeppesen, Kristensen, Nielsen, Humaidan, Dal Canto, Fadini, Schmidt, Ernst and Yding Andersen, 2012). (B) Data (markers) and simulation output (lines) of the relative expression of *FSH-receptor* (solid), *LH-receptor* (dashed), and aromatase (dotted) in the granulosa over developmental time. The data was recorded by (Jeppesen, Kristensen, Nielsen, Humaidan, Dal Canto, Fadini, Schmidt, Ernst and Yding Andersen, 2012) for follicles of different diameters. Data were converted from diameter into time by using the relation $d_F = 5mm + v_F t$. The vertical axis shows mRNA expression levels normalized to *Gapdh* expression. (C) The predicted maximal concentration of the LH receptor in theca (solid), granulosa (dash-dotted) and COC (dashed line) over developmental time.

**Figure 4** – Dynamics of spatiotemporal signalling in the follicle.

(A,B) Measured serum hormone concentrations of (A) FSH and (B) LH as reproduced (Brindle, Miller, Shofer, Klein, Soules and O'Connor, 2006). The solid lines show the interpolations used in the simulations. (C-F) Simulated concentration profiles of (C) FSH, (D) LH, (E) FSH receptor complex and (F) LH receptor complex relative to their signalling thresholds $K_F$ and $K_L$. The shaded areas indicate the standard deviation in the response when parameter values are sampled from a normal distribution with mean value as given in Table S1 and standard deviation $\sigma = 0.1$. Simulated (G) FSH receptor expression, and (H) LH receptor expression in the follicle at three time points: 2 (dotted), 6 (dashed), and 14 days (solid lines). Note that the simulations were carried out on a growing domain but are represented on a domain that is scaled such that all compartment sizes remain constant. The compartments on the horizontal axes indicate the different parts of the follicle, i.e. the theca (*T*), granulosa (*G*), COC (*COC*), and follicular fluid (*FF*). Note that in panels C-H only part of the follicular fluid domain is shown as the levels are constant within this domain.

**Figure 5** – Sensitivity of the Model Output to changes in Parameter Values and Initial Conditions. Sensitivity of the relative FSH concentration in the follicular fluid and the serum to 1% changes in the kinetic parameter values; only parameters with a sensitivity coefficient larger than 0.001 are included.

**Figure 6** – Predicted and Experimentally Confirmed FSH and LH gradients in the Follicle.
(A-B) The relative concentrations of (A) FSH, (B) hCG and LH in follicular fluid and serum for different serum concentrations. The data are based on measurements in the left and right ovary of 8 patients undergoing assisted reproduction. The trend-line shows the dependency of the ratio on the gonadotropin serum concentration. The shaded area marks the range of the serum concentrations used in the simulation based on the measured data in Fig 4A,B, the dotted line represents the serum concentration at the last time point (14 days). (C, D) The model predicts a gonadotropin concentration difference between the serum and the follicular fluid. The extent of the difference depends on the serum concentration. The dependency obtained in the model is similar to the one observed in the data; the trend line from the data is reproduced in panels C and D for ease of comparison. In the simulations constant serum levels were used over developmental time.

**Figure 7 -** Diffusion-based limits in granulosa cell layer thickness can explain scaling of follicle size with species weight. (A) The follicle surface area correlates with the weight of mammalian species (Table 1). (B) The rate of oestradiol production of the mature follicle scales with the surface area of the mature follice. Here only the volume of the antrum was increased; the thickness of the granulosa and the diameter of the COC was kept constant. (C-F) Receptor levels and granulosa width determine the extent of gonadotropin sequestration in the granulosa. The average FSH (black solid line), LH (black, broken line) and oestradiol (grey) concentration in the follicular fluid at day 14 (C) as the granulosa thickness is changed from its standard value denoted by 1, (D) as the *FSH-* and *LH-receptor* expression rates are changed from their standard value denoted by 1, (E) as both the granulosa thickness and the FSH serum levels are changed from their standard value denoted by 1, and (F) as both FSH serum levels and the *FSH-* and *LH-receptor* expression rates are changed from their standard value denoted by 1. Panels a and b were simulated on a constant 3D-domain of different radius as indicated. All other panels were simulated on growing domains.

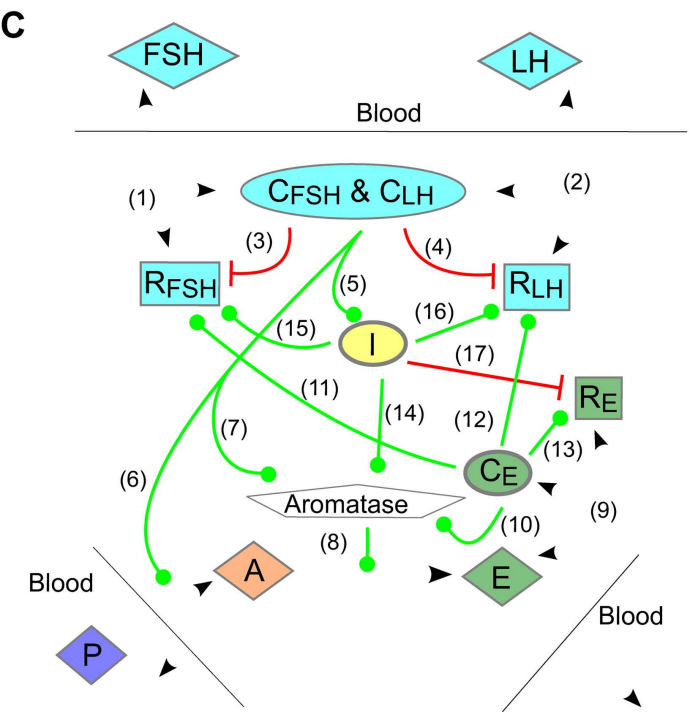

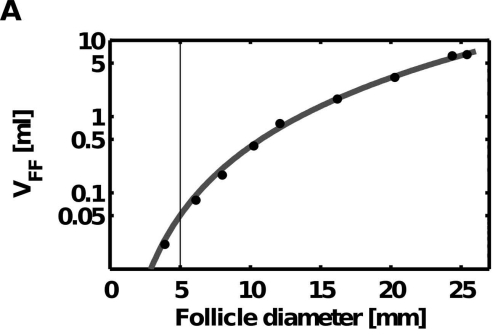
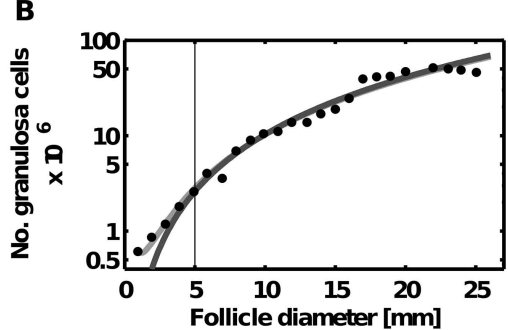
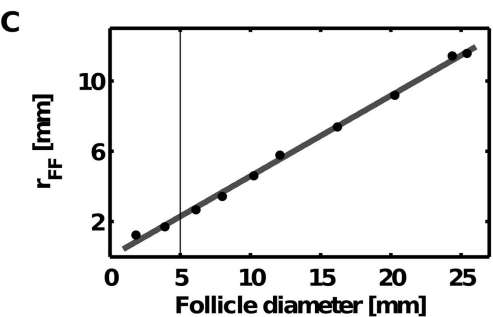
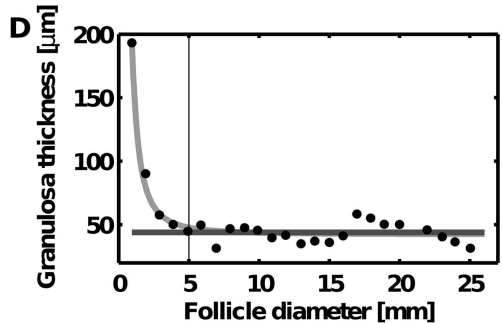
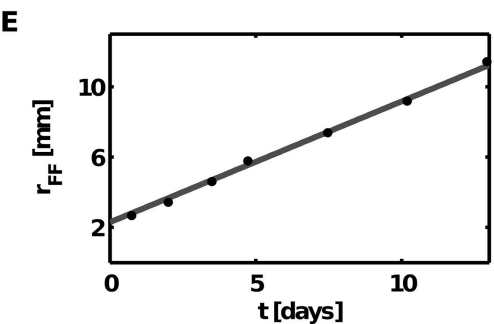
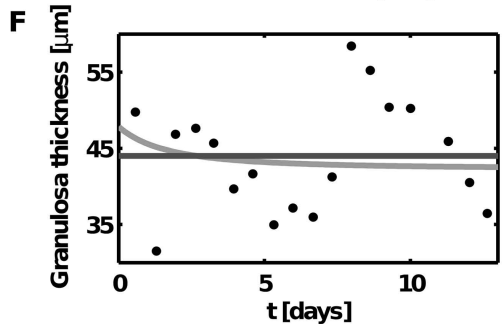
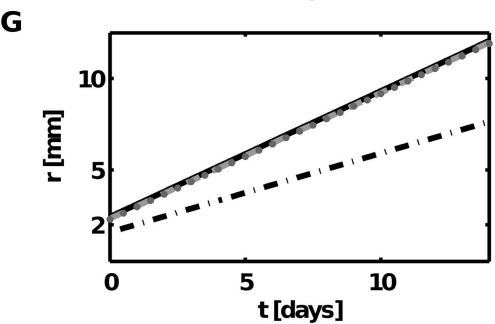
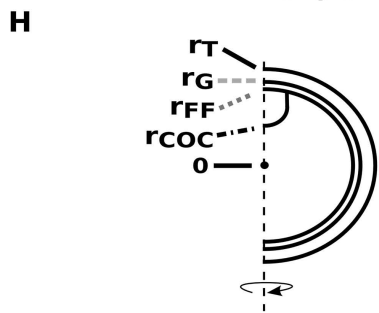

A

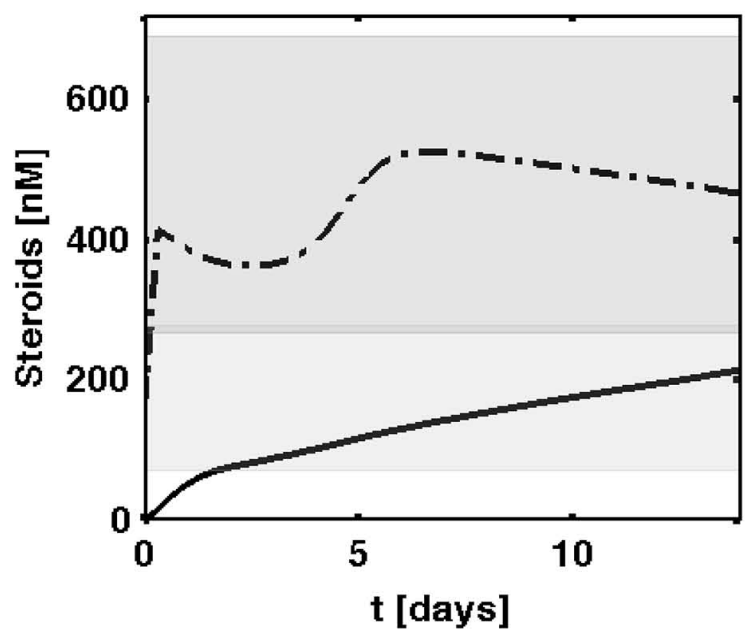

B

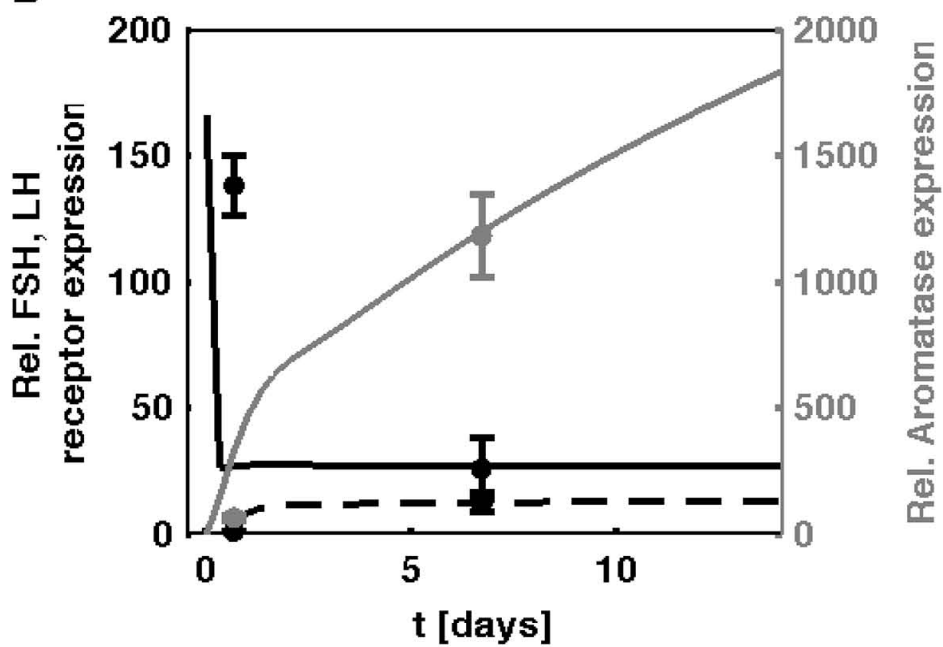

C

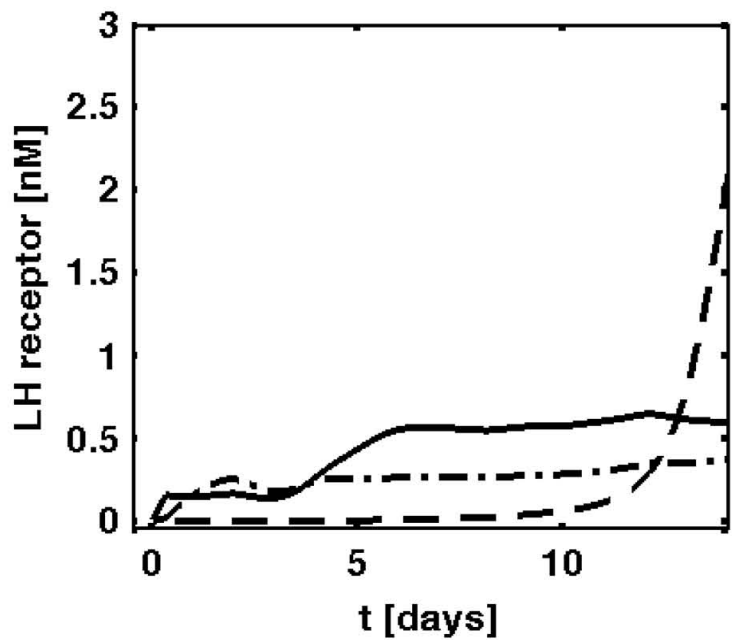

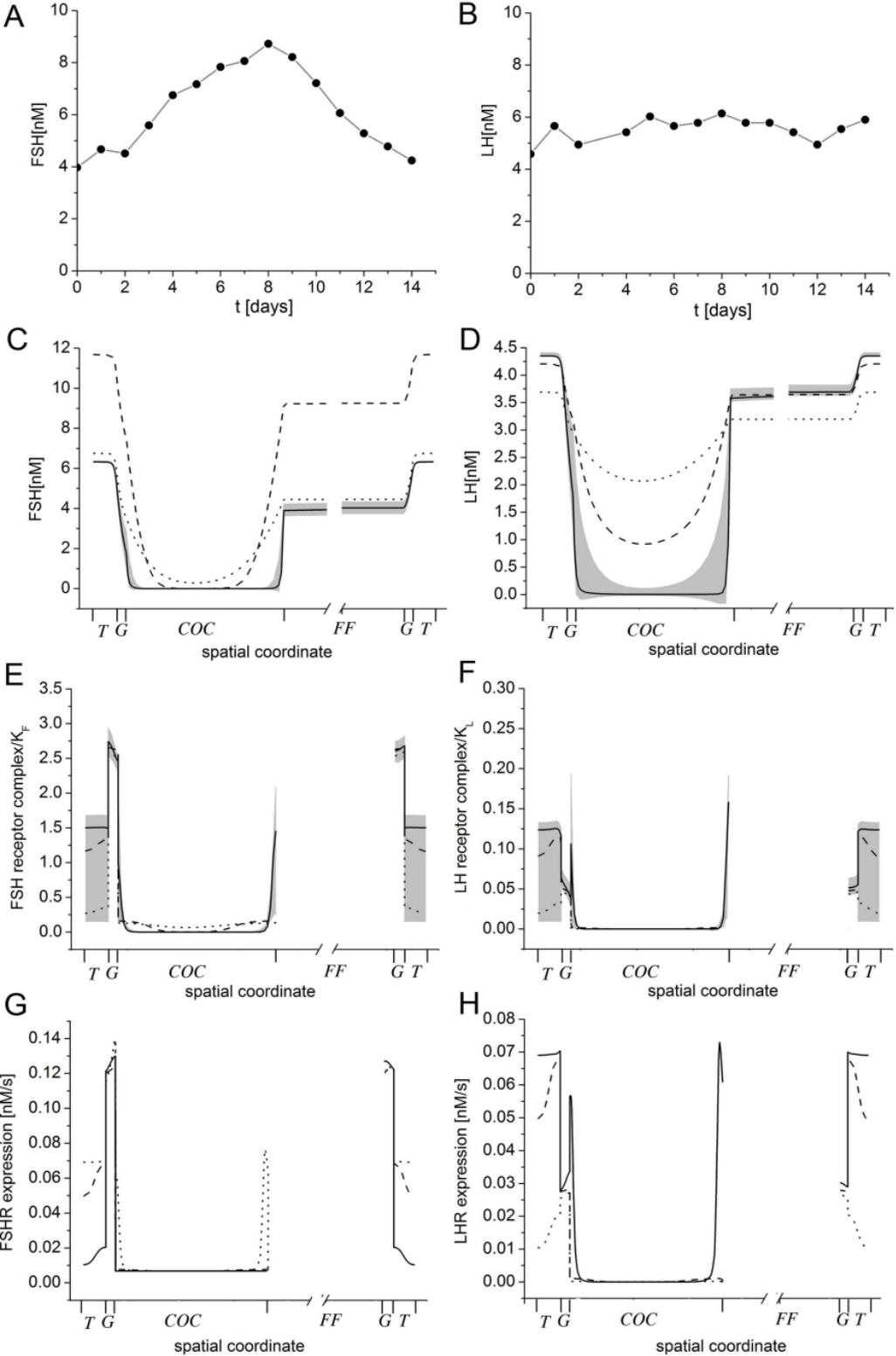

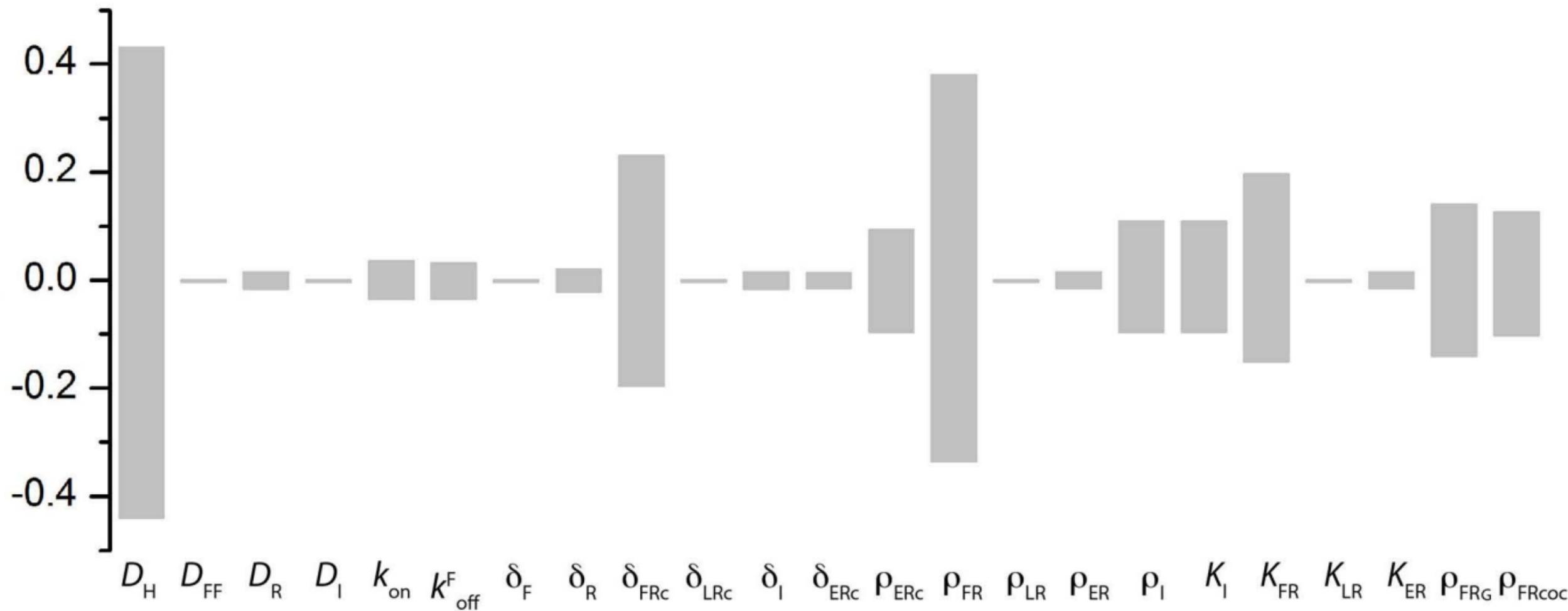

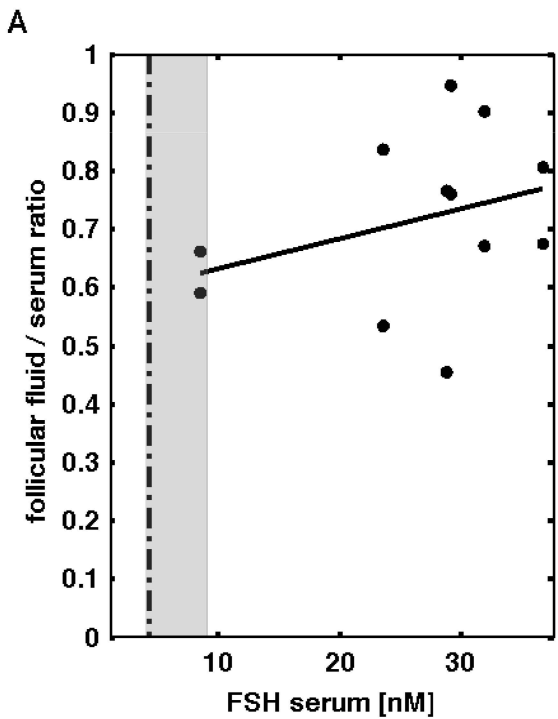 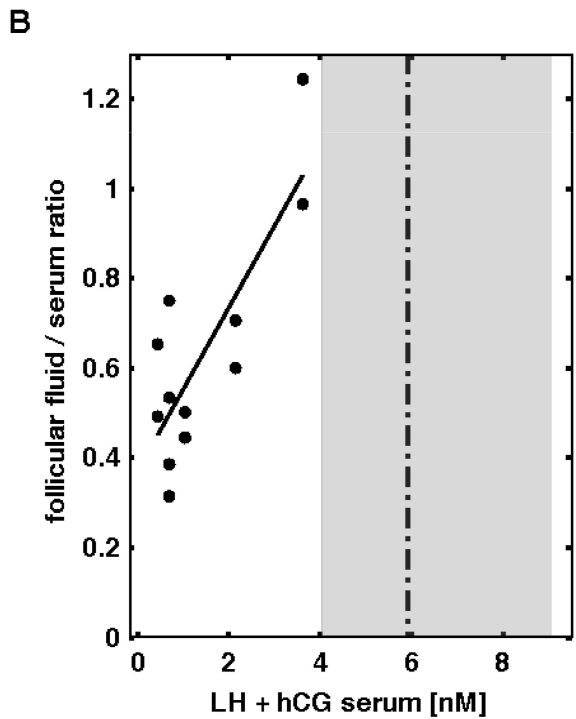
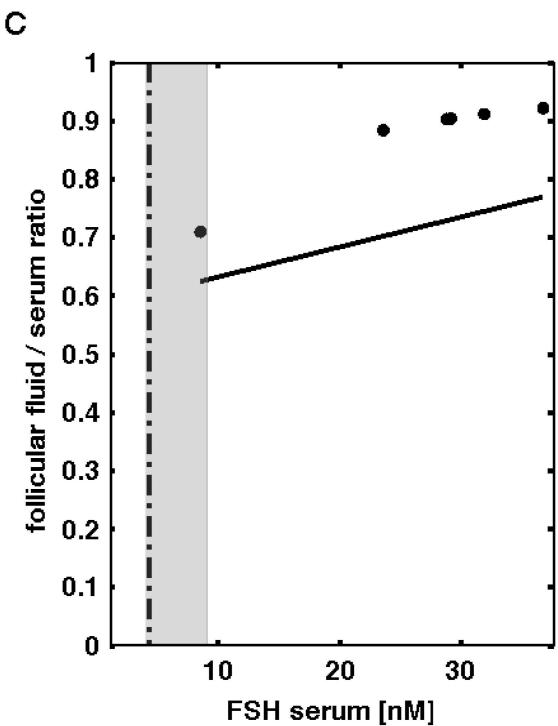 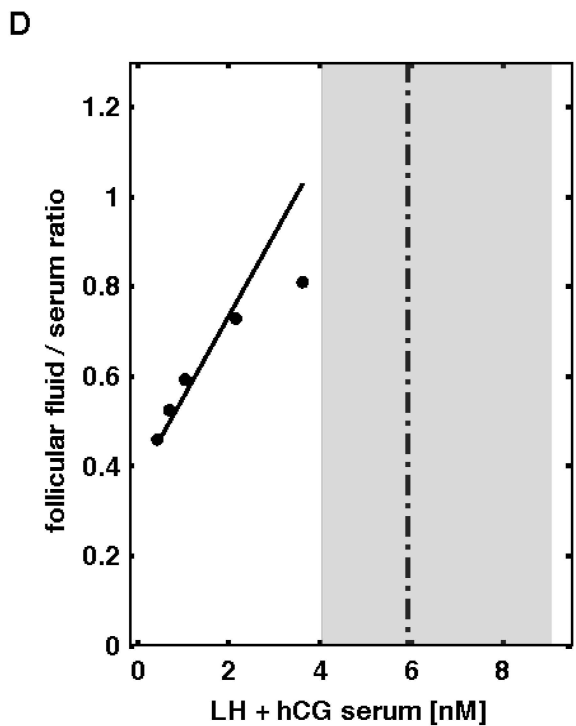

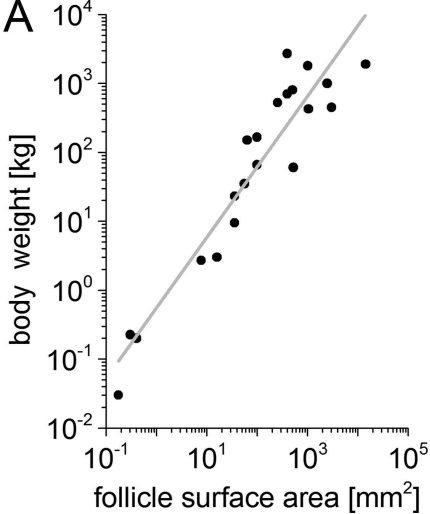
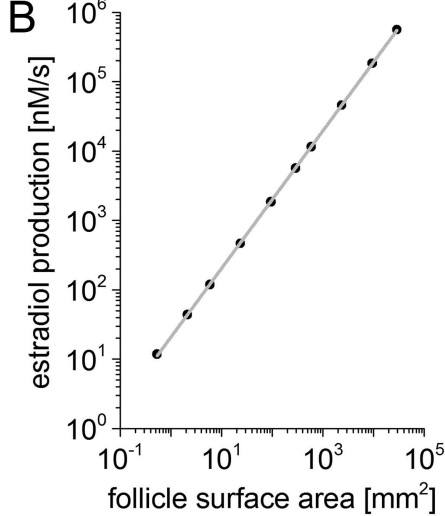
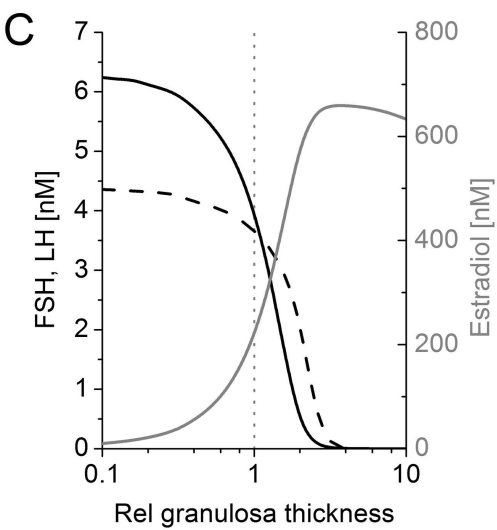
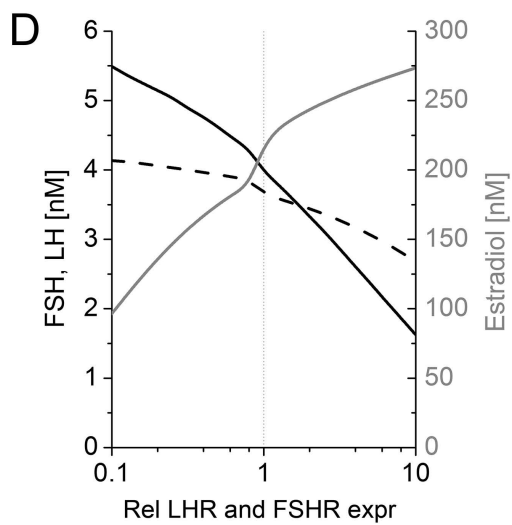
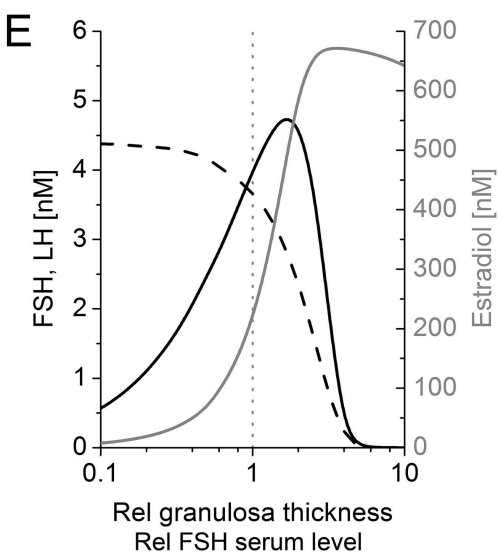
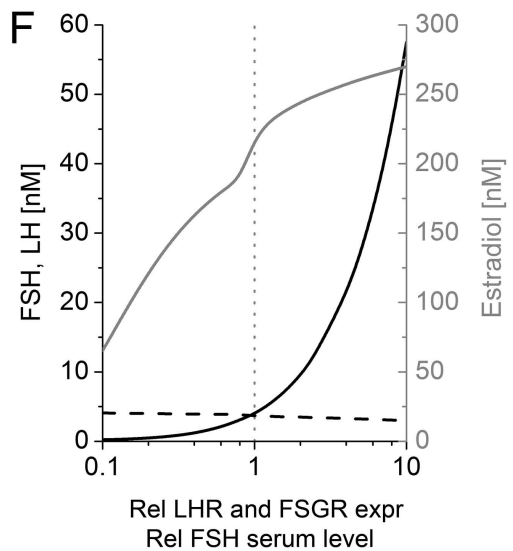

# Scaling of antral follicle size with species size as a result of diffusion-based limitations on ovarian granulosa thickness

## Supplementary Material

Mirjam Baechler[1,†], Denis Menshykau[1,2,†], Christian De Geyter[3], Dagmar Iber[1,2,*]

[1] Department for Biosystems Science and Engineering (D-BSSE), ETH Zurich, Switzerland
[2] Swiss Institute of Bioinformatics, Basel, Switzerland
[3] Division of Gynecological Endocrinology and Reproductive Medicine, Women's Hospital, University of Basel, Switzerland
† The authors contributed equally.
* Corresponding Author: Dagmar Iber (dagmar.iber@bsse.ethz.ch)

## The Model

The model is based on our previous model for bovine folliculogenesis [1] and is formulated as a set of isotropic advection-reaction-dispersion equations for a compound $c_i$ with diffusion coefficient $D_i$ and reaction terms $\mathcal{R}(c_i)$:

$$\partial_t c_i + \nabla \cdot (\boldsymbol{u} c_i) = D_i \nabla^2 c_i + \mathcal{R}(c_i) \tag{1}$$

where $\boldsymbol{u}$ denotes the external velocity field. The reaction terms $\mathcal{R}(c_i)$ are:

$$\mathcal{R}(F) = (\underbrace{\rho_F}_{\text{delivery}} \underbrace{-\Delta F}_{\text{removal}})\Theta \underbrace{-k_{on} F R_F + k_{off}^F C_F}_{\text{complex formation}} \underbrace{-\delta_F F}_{\text{decay}}$$

$$\mathcal{R}(L) = (\underbrace{\rho_L}_{\text{delivery}} \underbrace{-\Delta L}_{\text{removal}})\Theta \underbrace{-k_{on} L R_L + k_{off}^L C_L}_{\text{complex formation}} \underbrace{-\delta_L L}_{\text{decay}}$$

$$\mathcal{R}(A) = \underbrace{-\Delta A}_{\text{removal}}\Theta \underbrace{+\rho_A \Theta(1+\sigma_G)}_{\text{production}} \underbrace{-\delta_A A}_{\text{decay}} \underbrace{-\rho_E(\Gamma+\chi)\frac{A}{A+K_M}I(1+\sigma_E \sigma_G)(1+\sigma_G)}_{\text{catalytic decay}}$$

$$\mathcal{R}(E) = \underbrace{-\Delta E}_{\text{removal}}\Theta \underbrace{-k_{on} E R_E + k_{off}^E C_E}_{\text{complex formation}} \underbrace{-\delta_E E}_{\text{decay}} \underbrace{+\rho_E(\Gamma+\chi)\frac{A}{A+K_M}I(1+\sigma_E \sigma_G)(1+\sigma_G)}_{\text{catalytic production}}$$

$$\mathcal{R}(R_F) = \underbrace{\rho_{R_F}(\mathbf{1}-\sigma_{\mathbf{F}})(\mathbf{1}-\sigma_{\mathbf{L}})(\Gamma \vartheta_G + \chi \vartheta_{COC} + \sigma_I(1+\sigma_E))}_{\text{production}} \underbrace{-k_{on} F R_F + k_{off}^F C_F}_{\text{complex formation}} \underbrace{-\delta_R R_F}_{\text{decay}}$$

$$\mathcal{R}(R_L) = \underbrace{\rho_{R_L}(\mathbf{1}-\sigma_{\mathbf{F}})(\mathbf{1}-\sigma_{\mathbf{L}})\sigma_I(1+\sigma_E)}_{\text{production}} \underbrace{-k_{on} L R_L + k_{off}^L C_L}_{\text{complex formation}} \underbrace{-\delta_R R_L}_{\text{decay}}$$

$$\mathcal{R}(R_E) = \underbrace{\rho_{R_E}(1+\sigma_E)(1-\sigma_{I1})}_{\text{production}} \underbrace{-k_{on} E R_E + k_{off}^E C_E}_{\text{complex formation}} \underbrace{-\delta_R R_E}_{\text{decay}}$$

and

$$\begin{aligned}
\mathcal{R}(C_F) &= \underbrace{k_{on}FR_F - k_{off}^F C_F}_{\text{complex formation}} \underbrace{-\delta_{CF}C_F}_{\text{decay}} \\
\mathcal{R}(C_L) &= \underbrace{k_{on}LR_L - k_{off}^L C_L}_{\text{complex formation}} \underbrace{-\delta_{CL}C_L}_{\text{decay}} \\
\mathcal{R}(C_E) &= \underbrace{k_{on}ER_E - k_{off}^E C_E}_{\text{complex formation}} \underbrace{-\delta_{CE}C_E}_{\text{decay}} \\
\mathcal{R}(I) &= \underbrace{\rho_I \sigma_G}_{\text{production}} \underbrace{-\delta_I I}_{\text{decay}}
\end{aligned} \quad (2)$$

The reaction terms $\mathcal{R}_{c_i}$ are identical to those used in the bovine model [1], except for two new regulatory interactions (Fig. 1C A3, A4) that had to be added to the human model to account for a marked difference in the bovine and human gene expression data. Thus in human granulosa cells FSH receptor expression is high in small follicles (6mm) and subsequently decreases [2] while in bovine follicles FSH receptor expression is lower and increases over developmental time [3, 4]. To account for this difference we needed to introduce a previously neglected negative feedback of gonadotropin signaling on FSH and LH receptor expression. The bold terms in the equations for $\mathcal{R}_{R_F}$ and $\mathcal{R}_{R_L}$ reflect this negative feedback.

Note that the spatial restriction of reactions to one of the compartments (Fig. 1A,B) is incorporated by multiplication with Θ which is one in the theca and zero elsewhere, Γ which is one in the granulosa layer and zero elsewhere, or χ which is 1 in the COC and zero elsewhere.

The $\sigma$ terms indicate Hill functions that we use to describe such regulatory influences. To describe activating influences of a component $c_i$ we write

$$\sigma_i = \frac{c_i^{n_j}}{c_i^{n_i} + K_i^{n_i}}. \quad (3)$$

and we use $1 - \sigma_i$ to describe inhibitory impacts of $c_i$. Here $i$ denotes the component $i$. $K_i$ is the Hill constant which specifies the concentration of $c_i$ where half-maximal activity is observed, and the Hill coefficient $n_i$ defines the steepness of the response. We are using $n_i = 2$ throughout.

All other parameter values are listed in Table 1.

### Boundary Conditions

We use zero flux boundary conditions for all hormones, receptors and their complexes, i.e.

$$\nabla c_i = 0. \quad (4)$$

The receptors can diffuse slowly within a tissue but not between different cell layers, i.e. no flux between theca/granulosa, granulosa/COC, granulosa/fluid, COC/fluid. In this way it is ensured that receptors localize into/on their cells and are not present inside the follicular fluid. Gonadotropins and steroids can freely diffuse throughout the follicle. The diffusion constant in in the follicular fluid is set to a very high value ($D_{FF} = 0.1$ mm$^2$ s$^{-1}$) to account for the rapid mixing in the fluid as the women move.

## Initial Conditions

As initial conditions we use zero for the hormones, receptors and hormone-receptor complexes, because we want to study the mechanisms that result in the emergence of the characteristic gene expression patterns in the follicle, i.e.

$$\begin{aligned} F(0) &= L(0) = E(0) = A(0) = 0 \\ R_F(0) &= R_L(0) = R_E(0) = 0 \\ C_F(0) &= C_L(0) = C_E(0) = 0. \end{aligned} \tag{5}$$

The only exception is the initial concentration of the IGF-receptor complex. IGF-2 and the IGF type receptor are expressed in the theca already at the time of antrum formation [5], and the early presence of the IGF-receptor complex in the theca is important in the model to reproduce the experimentally observed early expression of LH receptors in the theca [3, 4]. To reproduce the measured LH receptor production rate in the early follicle we require

$$I(0) = 0.3\, K_I \Theta. \tag{6}$$

where $\Theta$ indicates the restriction to the theca and $K_I$ is the Hill constant for IGF-dependent regulatory processes.

# Tables

**Table 1. Parameter Values.** The table summarizes all parameter values used in the model along with the evidence. The first 26 parameter values are the same as in the bovine model [1] and almost entirely reflect directly measured values. The 7 further parameter values in the second block had to be altered to account for differences in the bovine and human data as discussed in the main text. The geometry parameters in the third block were directly derived from measured data as summarized in Figure 2. In addition to the parameters listed above we used
$F(0) = L(0) = E(0) = A(0) = R_i(0) = C_i(0) = 0$ as initial conditions to study the emergence of pattern from zero initial conditions. The serum levels of FSH and LH change over the menstrual cycle and $\rho_F(t)$ and $\rho_L(t)$ are set to the measured hormone concentrations [6] shown in Figure 4A,B times the blood flux rate $\Delta$ that determines how quickly the local serum concentration is replenished. The flux rate $\Delta$ is not known, but as long as $\Delta \geq 1$ s$^{-1}$ the flux does not impact on the predicted expression patterns in the follicle; we use $\Delta = 1$ s$^{-1}$. $\delta_{R_i}$ refers to the decay rates of all receptors, i.e. $i = \{F, L, E\}$.

| Source | Parameter | Simulation Value | Reference Value | References |
|---|---|---|---|---|
| From bovine model [1] | $D_H$ | $6.7 \times 10^{-5}$ mm$^2$ s$^{-1}$ | average $6.7 \times 10^{-5}$ mm$^2$ s$^{-1}$ for FSH | [7, 8] |
| | $D_S$ | $10^{-4}$ mm$^2$ s$^{-1}$ | small molecule Diff coeff $10^{-4}$ mm$^2$ s$^{-1}$ | [9, 10] |
| | $D_{C_L}$ | $2 \times 10^{-8}$ mm$^2$ s$^{-1}$ | $1.9 \pm 1 \times 10^{-8}$ mm$^2$ s$^{-1}$ | [11] |
| | $D_R$ | $10^{-7}$ mm$^2$ s$^{-1}$ | $10^{-9} - 5 \times 10^{-7}$ mm$^2$ s$^{-1}$ | [12–15] |
| | $D_I$ | $4 \times 10^{-9}$ mm$^2$ s$^{-1}$ | $3 - 5 \times 10^{-10}$ cm$^2$ s$^{-1}$ at 23 C; immobilization at 37 C | [12] |
| | $D_{FF}$ | $0.1$ mm$^2$ s$^{-1}$ | rapid mixing in follicular fluid, i.e. fast diffusion $D \gg 10^{-4}$ mm$^2$ s$^{-1}$ | [9, 10] |
| | $k_{on}$ | $10^{-3}$ nM$^{-1}$ s$^{-1}$ | standard value | [16] |
| | $k^F_{off}$ | $5 \times 10^{-4}$ s$^{-1}$ | average: $K_D = 5 \times 10^{-10} M^{-1}$ | [17–22] |
| | $k^L_{off}$ | $10^{-2}$ s$^{-1}$ | $K_D = 9$ nM | [23] |
| | $k^E_{off}$ | $10^{-4}$ s$^{-1}$ | $K_D = 0.1$ nM | [24] |
| | $\delta_F$ | $10^{-5}$ s$^{-1}$ | half-life of FSH in ovariectomized ewes is 20 h | [25] |
| | $\delta_L$ | $5 \times 10^{-4}$ s$^{-1}$ | half-life of LH in ovariectomized rats is 23 minutes | [26] |
| | $\delta_{R_i} = \delta_{C_E} = \delta_E = \delta_A$ | $3 \times 10^{-5}$ s$^{-1}$ | half-life 1-6 hours | [27–31] |
| | $\delta_{C_L}$ | $7.5 \times 10^{-4}$ s$^{-1}$ | half-life LH/hCG-bound receptor 17 min | [26, 27, 32], |
| | $\delta_{C_F}$ | $6.4 \times 10^{-5}$ s$^{-1}$ | half-life of FSH is 3 hours | [25] |
| | $\delta_I$ | $10^{-6}$ s$^{-1}$ | half-life 6 – 10 hours, i.e $\sim \delta = 10^{-6}$ s$^{-1}$ | [33] |
| | $\rho_{R_F} = \rho_{R_L}$ | $500 \times 0.5$ pM s$^{-1}$ | 21 receptors/cell/min = 0.5 pM s$^{-1}$, surface receptor | [27, 34–37] |
| | $\rho_E$ | $0.06$ s$^{-1}$ | $k_{cat} = 0.06$ s$^{-1}$ | [38, 39] |
| | $\rho_{R_E}$ | $1.25$ pM s$^{-1}$ | 45000 estrogen receptors are detected per cell | [40] |
| | $K_M$ | 44 nM | $K_M = 44$ nM | [38, 39] |
| | $K_E$ | 35 nM | reproduce bovine FSH and LH receptor expression kinetics | [1, 3, 4] |
| | $I(0)$ | $0.3 K_I \Theta$ | reproduce bovine LH receptor expression kinetics | [1, 3, 4] |
| Human data | $\rho_I$ | $13.5 K_I \delta_I$ | reproduce measured granulosa LH receptor numbers ($\leq 3$ nM) | [34] |
| | $\rho_A$ | $\Delta \times 400$ nM | reproduce androgen concentration in follicular fluid, Fig. 3A | [2] |
| | $K_I$ | 35 nM | reproduce estradiol concentration in follicular fluid, Fig. 3A | [2] |
| | $K_F = K_L$ | $0.45 \, \vartheta \times 500$ | reproduce steepness of expression kinetics in Fig. 3B | [2] |
| | $\vartheta_G$ | 3 | reproduce early FSH receptor expression rate in Fig. 3B | [2] |
| | $\vartheta_{COC}$ | $0.009 \times \vartheta_G$ | limit total receptor expression in COC to physiological range | [34] |
| Geometry | $r_{FF}(0)$ | 2.292 mm | ordinate intercept in Fig. 2E | [41] |
| | $r_G(0) - r_{FF}(0)$ | 44 $\mu$m | constant fit of data points in Fig. 2F | [41] |
| | $r_T(0) - r_G(0)$ | 100 $\mu$m | $\sim 50 - 100$ $\mu$m | [42] |
| | $r_{COC}(0)$ | 1.942 mm | ordinate intercept in Fig. 2G | [41] |
| | $v_{FF}$ | $0.8 \times 10^{-5}$ mm s$^{-1}$ | fit of data points in Fig. 2C | [41] |
| | $v_F$ | $1.74 \times 10^{-5}$ mm s$^{-1}$ | $\sim 5 - 25$ mm in 14 days | [43] |